\documentclass[useAMS,usenatbib,usegraphicx]{mn2e}
\title [On the detection of Trojan planets in binaries]
{Eclipse timing variations to detect possible Trojan planets in binary systems}
\author[R. Schwarz, \'A. Bazs\'o, B. Funk and R. Zechner]
{R. Schwarz\thanks{E-mail:schwarz@astro.univie.ac.at}, \'A. Bazs\'o, B. Funk and R. Zechner\\
Institute for Astronomy, University of Vienna, A-1180 Vienna, 
T\"urkenschanzstrasse 17, Austria\\ 
}

\begin{document}

\date{Accepted 1988 December 15. Received 1988 December 14; in original 
form 1988 October 11}

\pagerange{\pageref{firstpage}--\pageref{lastpage}} \pubyear{2002}
\maketitle
\label{firstpage}

\begin{abstract}
This paper is devoted to study the circumstances favourable to detect Trojan 
planets in close binary-star-systems by the help of eclipse timing variations 
(ETVs). 
To determine the probability of the detection of such variations with ground 
based telescopes and space telescopes (like former missions CoRoT and Kepler 
and future space missions like Plato, Tess and Cheops), we investigated the 
dynamics of binary star systems with a planet in tadpole motion.
We did numerical simulations by using the full three-body problem 
as dynamical model. The stability and the ETVs are investigated by computing 
stability/ETV maps for different masses of the secondary star and the Trojan 
planet. 
In addition we changed the eccentricity of the possible Trojan planet. 
By the help of the libration amplitude $\sigma$ we could show whether or not 
all stable objects are moving in tadpole orbits. 
We can conclude that many amplitudes of ETVs are large enough to detect 
Earth-like Trojan planets in binary star systems. As an application, we
prepared a list of possible candidates. 
\end{abstract}

\begin{keywords}
celestial mechanics -- methods: numerical -- (stars): planetary systems -- 
(stars):binaries: general
\end{keywords}

\section{Introduction}
\label{intro}

The first extra solar planet was discovered in the early 1990s by 
\citet{w}. Today the statistics of the observations showed that the 
architecture of our solar system seems to be unique compared with exoplanetary 
systems. At the moment we know about 1900 exoplanets in more than 1200 
planetary systems, among them more than 100 exoplanets are in binary-star 
systems and 20 are in multiple-star systems. The data of all planets are 
collected in the Exoplanet-catalogue maintained by J. 
Schneider\footnote{http://exoplanet.eu}; 
whereas the binary and multiple-star systems can be found separately in the 
binary catalogue of 
exoplanets\footnote{http://www.univie.ac.at/adg/schwarz/multiple.html} 
maintained by R. Schwarz.
For nearby Sun-like stars, approximately 70 percent are confirmed to be double 
or multiple systems ($\sim 75$\% for O-B stars, \citet{versch}, \citet{mason}; 
$\sim 67$\% for G-M stars, \citet{mayor}). 
Statistics of solar-type dwarfs were studied by \citet{toko14} with a distance-limited
sample of 4847 targets. A field population was found of about 54\% for single stars,
33\% binary stars, 8\% triple systems, 4\% for quadrupole systems, 1\% for systems $N>4$.

According to the work of \citet{rabl} one can distinguish three types of 
planetary orbits in a binary star system:
\begin{enumerate}
\item S-Type, where the planet orbits one of the two stars;
\item P-Type, where the planet orbits the entire binary;
\item T-Type: a planet may orbit close to one of the two equilibrium points 
$L_4$ and $L_5$ (Trojan planets).
\end{enumerate}

From the dynamical point of view the binary star systems are particularly 
interesting. For the circular restricted three-body problem (CR3BP) it has 
been shown by \citet{gascheau} and \citet{routh} that only for mass ratios in 
the interval $0\leq\mu\le\mu_{crit}$ the librational motion is linearly stable 
(here $\mu=m_2/(m_1+m_2)$, and $\mu_{crit}=(1-\sqrt{23/27})/2\approx0.0385)$.
However, former studies of \cite{erdi09} and \cite{sic}
could show that there exist stable orbits beyond Gascheau’s value ($\mu_{crit}$), 
in the planar case. Whereas, \cite{akos} could also show that in the spatial 
restricted three-body problem. However, even the extended stability region 
limits the possibility to find Trojans in double stars, because the necessary 
mass ratios are relatively rare.

\begin{figure}
\centerline
{\includegraphics[width=7.1cm,angle=0]{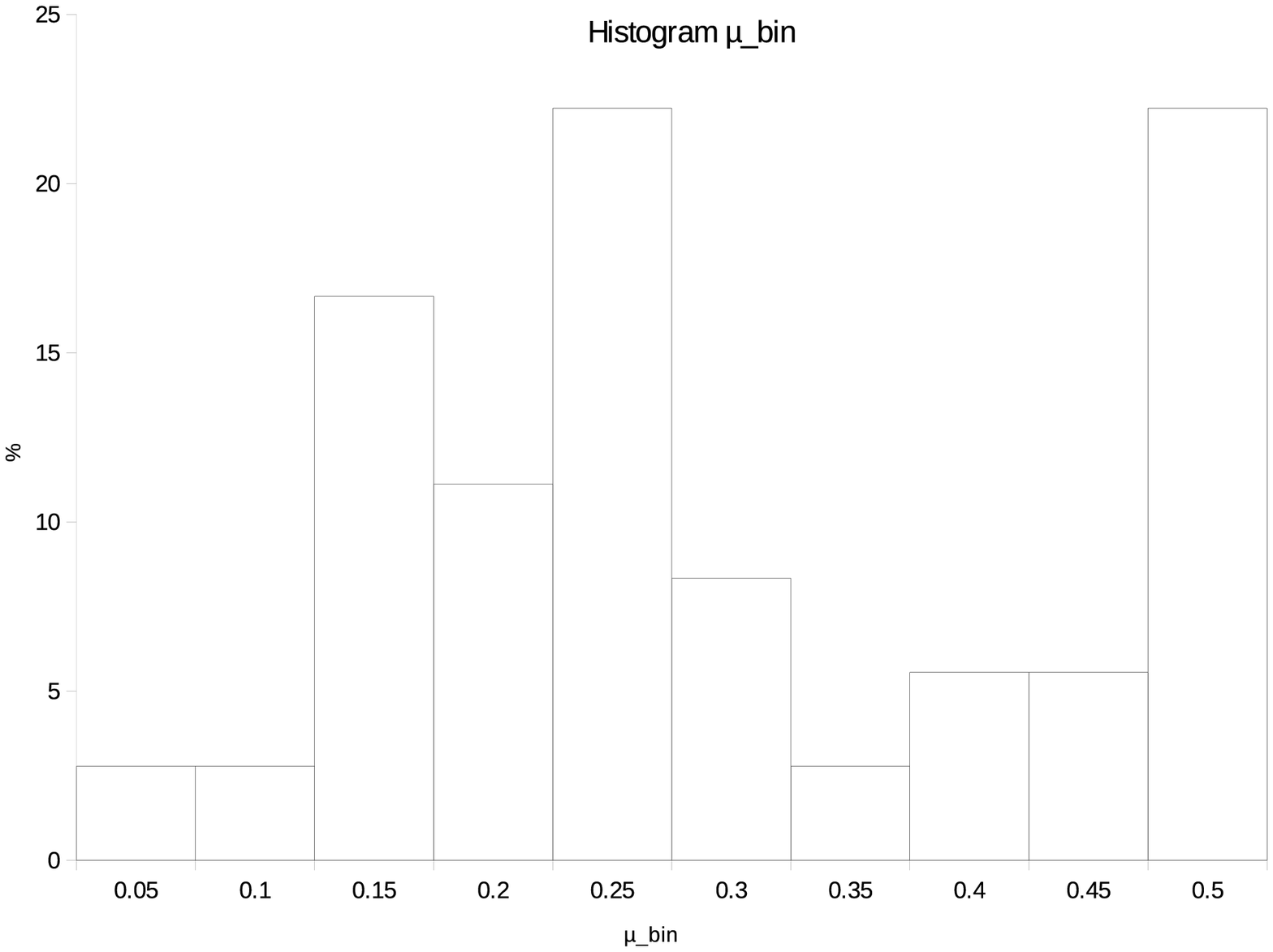}}
\caption{The Histogram present the mass ratio $\mu_{bin}$ of all binary star 
systems with exoplanets, taken from binary catalogue of exoplanets 
(http://www.univie.ac.at/adg/schwarz/multiple.html).}
\label{fig1}
\end{figure}

Most observations of planets in binaries are focused on $\mu\approx 0.5$ 
(stars have similar masses) and are restricted sun like stars. 
Nevertheless the dynamical study of \citet{sch09} could show with a few real 
binary systems that the T-Type configuration may be not only of theoretical interest.
Therefore we show a distribution (Fig.~\ref{fig1}) of the mass ratio of all 
detected exoplanets in binaries. The most common mass ratios are 0.25 and 0.5. 
The frequency of the interesting mass ratios 
($\mu_{crit}$) for the Trojan configuration is around 3 percent. When we 
compare Fig.~\ref{fig1} with the statistics of the nearby Sun-like stars the 
value is larger, it is around 10 percent~\citep{duq91}. However, if one considers
also systems with a brown dwarf as companion the number of possible candidates 
would be larger (as shown in our candidate list Tab.~\ref{tab3}), due to their 
substellar masses.

Our work was motivated by the huge number of Trojans which were found in the 
solar system around different planets (Earth, Mars, Jupiter, Uranus and 
Neptune). 
The possibility of Trojan planets in exoplanetary 
systems was discussed in several dynamical investigations like e.g. 
\cite{nauen}, \citet{lc}, \cite{erdi05}, \cite{dvorak04}, and \cite{sch07}. 
Theoretical studies predict that Trojans are a by-product 
of planet formation and evolution, see the hydrodynamic simulations of a 
protoplanetary disk, like in the work of \citet{chiangl} 
and \citet{beauge}. Whereas \citet{pierens} studied the orbital evolution 
of co-orbital planets embedded in a protoplanetary disc.
The investigation on the possibility of a migration-induced resonance locking 
in systems containing three planets, namely an Earth-like planet, a super-Earth and 
a gas giant with one Jupiter mass by \citet{epod} found that the super-Earth 
planet captures the Earth-like planet into co-orbital motion temporarily.
In case of binary systems the formation or capture scenario for possible Trojan
planets are not yet investigated.

\citet{fordg} and \citet{fordh} examined the sensitivity of transit timing 
variations (TTVs) for a possible detection of Trojan companions. They 
demonstrated that this method offers the potential to detect terrestrial-mass 
Trojans using existing ground-based observatories.
\citet{janson} reported a systematic search for extrasolar Trojan companions
to 2244 known Kepler Objects, but no Trojan candidates were found.
In a theoretical work of~\citet{vn}, they studied the TTVs for planets in the 
horseshoe regime by using semi-analytical models. 

This paper is divided into two parts: the first part investigates
the stability of Trojan planets in binary systems, whereas the second part 
is devoted to the possible detection of Trojan planets by the help of eclipse 
timing variations (ETV).

\section{Numerical setup}
\label{setup}

\subsection{Models}
We studied the planar full three-body problem (3BP) with numerical integrators.
In this problem three finite bodies, the primaries 
($m_1=primary$ $star$, $m_2=secondary$ $star$) revolve about their common 
center of mass, starting on circular orbits ($e_2=0$), and a third body $m_{Tro}=Trojan$ 
moves under their gravitational influence in the same orbit as $m_2$ close to 
the equilibrium point $L_4$ 60 degree ahead of the secondary star.
\par
We have regarded all the celestial bodies involved as point masses and 
integrated the dimensionless (the semi-major axis set to one) equations of 
motion for an integration time up to $T_c=10^4$ and $10^6$ periods of the secondary star 
for the stability maps and $T_c=10^7$ periods for special cases shown in chapter 
\ref{stab}.
For our simulations we used the Lie-method with an automatic step-size control 
to solve the equations of motion \citep{hanslmeier,licht,eggl}.

\subsection{Initial conditions} 
The computations were accomplished only in the vicinity of $L_4$ ($L_5$ is 
symmetric in the 3BP). 
For the calculations of the stability maps we varied the eccentricities of the 
Trojans ($e_{Tro}$=Trojan planet) from $e_{Tro}=0.01$ up to $e_{Tro}=0.5$. 
In addition, we varied the mass of the secondary object from 
$m_2=10^{-3}M_{sun}$ ($1 M_{sun}$ corresponds to 1 Solar mass) until the 
end of the stability border with a map size of 235~x~40 values 
($m_2$ x $e_{Tro}$). The lower limit of $m_2$ presents planets of about
2 Jupiter masses as shown in our stability investigations. In case of the ETV's we started
our investigations at masses equal to substellar objects (approximately 13 Jupiter masses).  
The stability border changes because we also calculated 
the stability maps for different masses of the host star $m_1$=1, 2 and 
3~$M_{sun}$. 
For $m_1=1~M_{sun}$ the stability border lies at 
$m_2=0.045$ $M_{sun}$, for $m_1=2~M_{sun}$ at $m_2=0.087$ $M_{sun}$ and for 
$m_1=3~M_{sun}$ at $m_2=0.130$ $M_{sun}$.

To get a good estimation about occurring perturbations on the secondary star
(to measure ETVs) we used Trojan planets with different sizes: Mars, Earth, Neptune
and Jupiter, which corresponds to $10^{-3}$ $M_{sun}$.

\subsection{Methods} 
For the analysis of the orbit we used the method of the maximum 
eccentricity $e_{max}$. In former studies we found a good agreement with 
chaos indicators \citep[as shown in ][]{akos} like the Lyapunov characteristic 
indicator (LCI), like in ~\cite{sch07a}. 
The $e_{max}$ method uses as an indication of stability a straightforward 
check based on the maximum value of the Trojans eccentricity reached during
the total integration time. If the Trojan's orbit becomes parabolic 
($e_{max} \ge 1$) the system is considered to be unstable.  
The $e_{max}$ is defined as follows:

\begin{equation}
\mathrm{e_{max}} = {\max_{t \le T_{c}}(e)}.
\end{equation}

\par
To analyse the orbital behaviour of the third body $m_3$ (e.g. tadpole, 
horseshoe, or satellite), we used the amplitude of libration, as done in the
work of \citet{freist} and \cite{sch13}.

The classification of the orbit was done by checking the libration 
amplitude $\sigma$ which is defined as the difference between the mean 
longitude of the Trojan and the secondary star ($\lambda_{Tro} -\lambda_2$). 
$\lambda_{Tro}$, $\lambda_2$ are given by $\lambda_{Tro}=\varpi+M$, 
$\lambda_2=\varpi_2+M_2$ were $\varpi$, $\varpi_2$ are the longitudes of 
ascending node of the Trojan body and of the $m_2$ and $M_{Tro}$, $M_2$ 
are the mean anomaly of the third body respectively of the secondary star. 
As for the $e_{max}$ we determined the maximum libration amplitude 
$\sigma_{max}$ for the total integration time:

\begin{equation}
\mathrm{\sigma_{max}} = {\max_{t \le T_{c}}(\sigma)}.
\end{equation} 
 
In the case of a large libration amplitude the Trojan may become
a horseshoe orbit or may be ejected out of the Trojan region after a
close approach with the secondary body. The value of $\sigma_{max}$
should be smaller than $180^{\circ}$ for tadpole orbits.  
\par
As indicated by the results of the stability analysis, an exoplanet in a T-type 
orbit may be stable in the vicinity of a binary star system. The gravitational 
perturbation of this planet may affect the motions of the two
stars and cause their orbits to deviate from Keplerian. In
an eclipsing binary, these deviations result in variations in
the time and duration of the eclipse.

We determined the ETVs by calculating the perturbed case, where the planet induced 
constant rate of apsidal precession is removed by a linear fit. We also took into account
the long-term effects caused by the binaries motion around the system center of mass.

\section{Stability}
\label{stab}

\begin{figure}
%\centerline
{\includegraphics[width=8.0cm,angle=0]{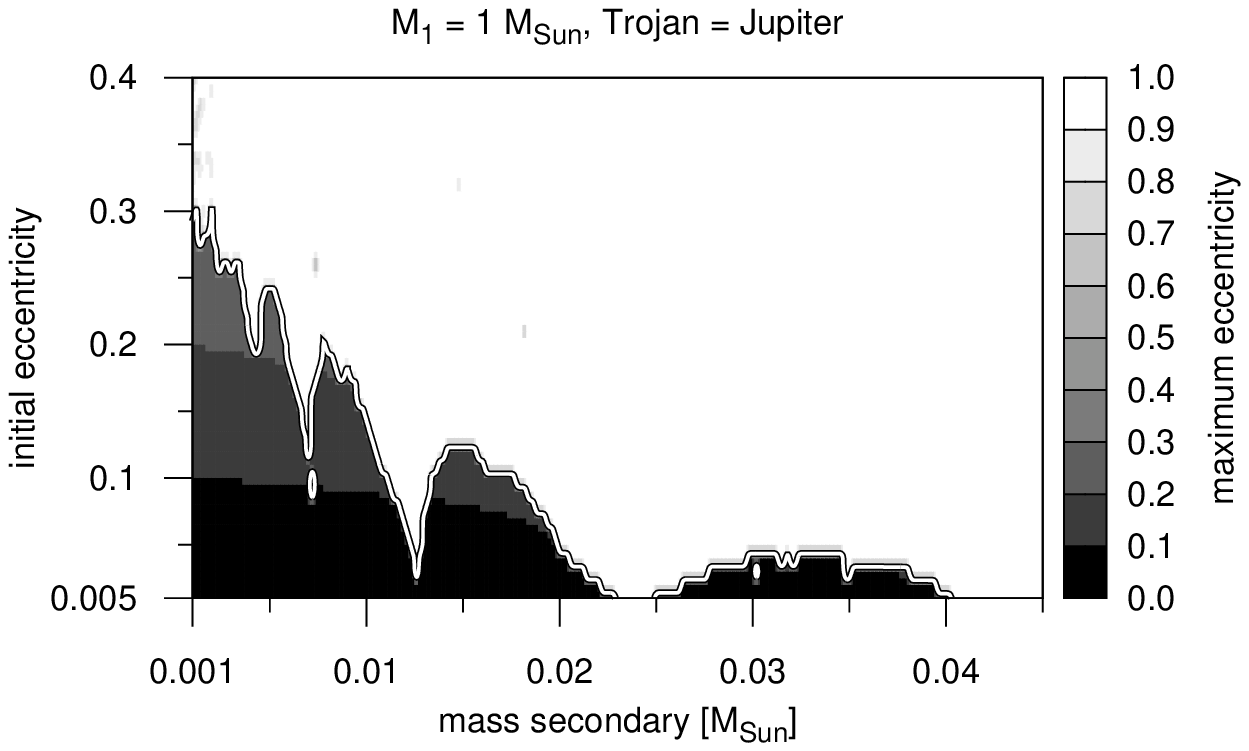}}
{\includegraphics[width=8.0cm,angle=0]{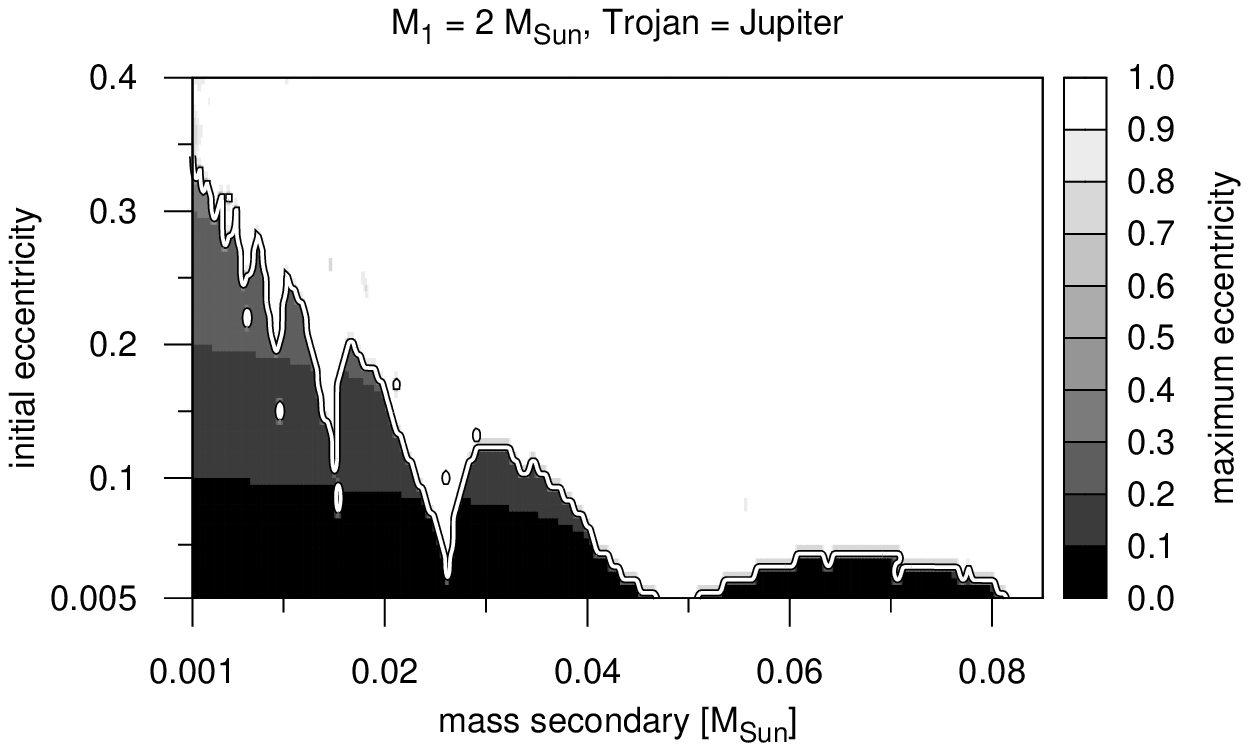}}
{\includegraphics[width=8.0cm,angle=0]{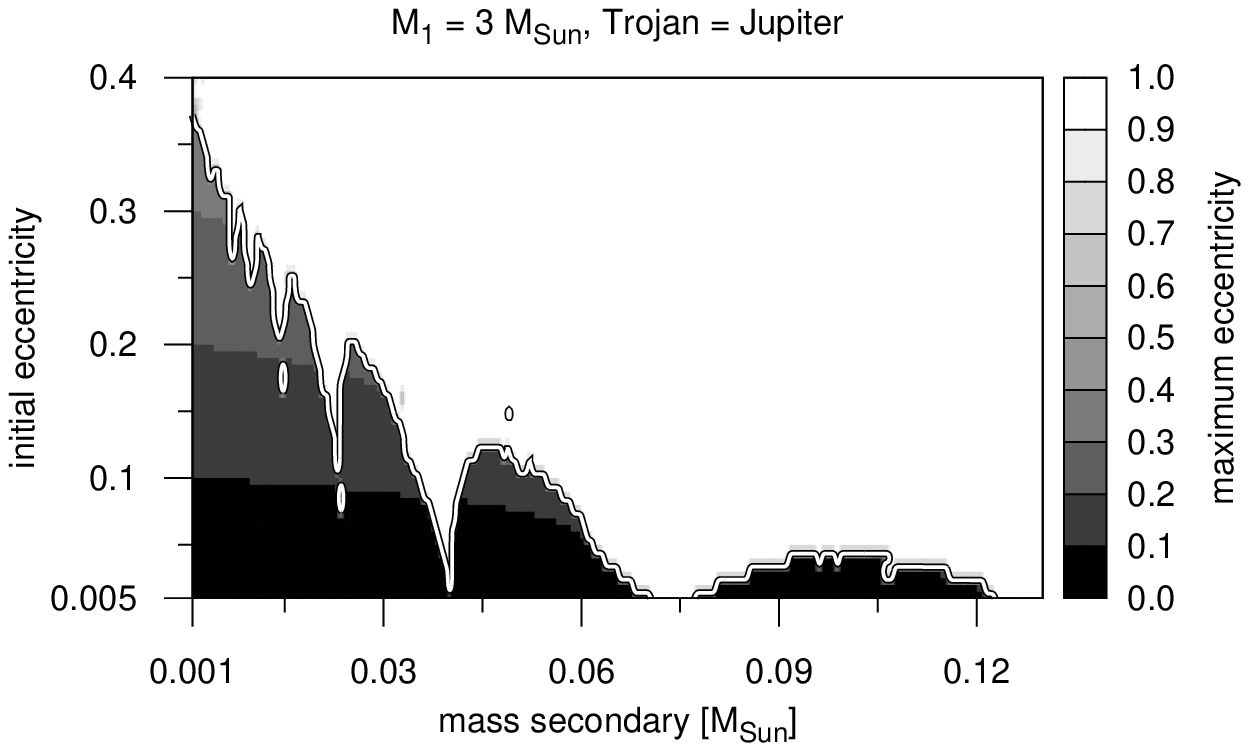}}
\caption{Stability map for $L_4$ in the planar ($i=0^{\circ}$) 
three-body problem for a Trojan planet with 1~$M_{Jup}$, for 1~$M_{sun}$ 
(upper graph) 2~$M_{sun}$ (middle graph) and  3~$M_{sun}$ (lower graph). 
$L_4$ is stable in the light, unstable in the dark region for the $e_{max}$ 
and the LCI. The grey scale presents the values of $e_{max}$, where the dark 
regions represent small values $e_{max}$ (stable orbits) and light regions 
depicts large values of  $e_{max}$ (unstable orbits).}
\label{fig2}
\end{figure}

First of all we investigated the stability of the Lagrangian point
$L_4$ in the planar three-body problem to determine the stability
limits and to investigate whether the orbits are still in Trojan motion or not, 
by the help of $\sigma_{max}$. Our studies showed that all stable orbits stay in 
tadpole motion.
In addition we used different masses of the primary and secondary star, which 
might be useful for future observations of different stars. We computed the 
stability maps by changing the mass of the secondary and the eccentricity of 
the Trojan.
An example is shown in Fig.~\ref{fig2}, for the 3 different masses of the 
primary star (1, 2 and 3~$M_{sun}$).

\begin{figure}
\centerline
{\includegraphics[width=8.0cm,angle=0]{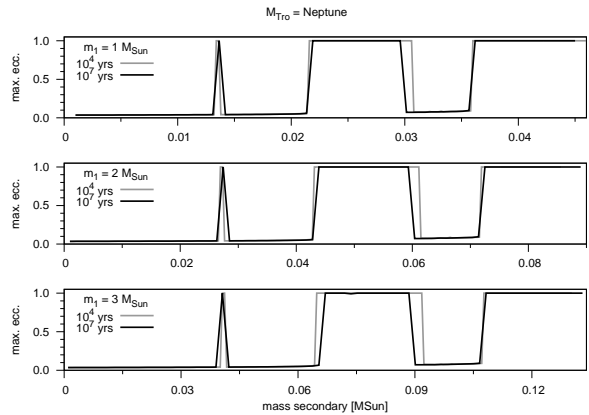}}
\caption{Cuts of the stability maps for a mass of the Trojan body of 
$M_{Tro}=1 M_{Neptune}$ for low eccentricity ($e_{Tro}=0.05$ and an integration 
time $T_c=10^4$ compared with $T_c=10^7$ periods.}
\label{fig3}
\end{figure}

As one can see the stability maps do not start at zero (x and y axes in 
Fig.~\ref{fig2}). 
That means the initial mass of the secondary body is 0.001~$M_{sun} \sim M_{Jup}$ 
and $e_{Tro}=0.005$. However, we focus on the 
stability region from $m_2 \ge 10 M_{Jup}$, because lower masses 
(planet-like masses) of the secondary are not interesting for 
binary (like) star-systems and initially circular orbits ($e_{Tro}=0$) 
do not produce detectable ETV signals. An example of the stability 
maps is given in Fig.~\ref{fig2}, which is divided into 3 different graphs, 
showing different masses of the primary star ($m_1$).
Whereas, the upper graph depicts the stability map for $m_1=1 M_{sun}$, the 
middle one depicts $m_1=2 M_{sun}$ and the lower one $m_1=3 M_{sun}$. We can 
conclude that the number of stable orbits do not differ much, as summarized in 
Tab.~\ref{tab1}, where we also show the studies for different masses
of the Trojan planet (from Earth mass up to Jupiter mass).
In principle the number of stable orbits do not change very much for the
different masses of the Trojan. 
Nevertheless, the shape of the stable region (gaps) are changing when we 
change the mass of the secondary star. The gaps are mainly caused by secondary 
resonances which were investigated in the work of \cite{sch12a} for large mass 
ratios and in \cite{sch13} for low mass ratios. In addition, one can see that 
the Trojan body can have larger initial eccentricity for larger mass of the 
primary star $m_1$, this is especially well visible for small masses of $m_2$. 
The maximum stable eccentricity $e_{ini}=0.245$ is valid for $m_2=0.01$ (which 
corresponds to a binary-like star system with a brown dwarf) and for different 
masses of central star ($m_1$) $M_{Tro}=1~M_{Jup}$.
$\mu_{crit}=0.04060$ for $1~M_{Sun}$, $\mu_{crit}=0.08185$ for $2~M_{Sun}$ and 
$\mu_{crit}=0.12365$ for $3~M_{Sun}$.
Also the limit of the stable region increases with a more massive central star: 
for $m_1=1 M_{sun}$ the limit amounts $m_2=0.041$, for $m_1=2 M_{sun}$ the limit 
amounts $m_2=0.082$ and for $m_1=M_{sun}$ we have the largest value $m_2=0.12$. 
The stability limit changes because of the different mass ratio 
$\mu=m_2/(m_1+m_2)$ as we know from the circular restricted three-body problem.
In addition, we made long term integrations, represented in Fig.~\ref{fig3} by 
the help of cuts of the stability maps shown in Fig.~\ref{fig2} 
(cuts for $e_{Tro}=0.035$).
The cuts were made for the different masses of the planets, as shown in the 
chapter of the initial conditions. 
We presented only the case of $M_{Tro}=1 M_{Neptune}$, because the other cases 
are similar. 
The goal of these cuts was to see whether the stability-island for high
mass ratios is still stable for a long time, which is the case.

\begin{table}
\centering
 \caption{Number of stable orbits of the stability maps for $T_c=10^6$ periods 
and for different masses of the primary body (central star) and third body 
(Trojan planet) presented in the second column starting with Earth mass 
($m_{Tro}=0.30049\cdot10^{-5}M_{sun}$), Neptune mass 
($m_{Tro}=0.51684\cdot10^{-4}M_{sun}$), and Jupiter mass 
($m_{Tro}=0.9547\cdot10^{-3}M_{sun}$).}
  \begin{tabular}{llll}
  \hline
mass of  & mass of   & number of & percent of\\
the primary & the Trojan & stable orbits& stable orbits\\
$[M_{sun}]$&body&&\\
\hline
        1 & Earth &   734 &   21.839\\
        1 & Neptune &   729 &   21.690\\
        1 & Jupiter &   639 &   19.012\\
\hline
        2 & Earth &  1523 &   22.947\\
        2 & Neptune &  1517 &   22.857\\
        2 & Jupiter &  1404 &   21.154\\
\hline
        3 & Earth &  2316 &   23.363\\
        3 & Neptune &  2305 &   23.252\\
        3 & Jupiter &  2168 &   21.870\\
  \hline
\end{tabular}
\label{tab1}
\end{table}

\section{Eclipse timing variations}
\label{res}

When we are looking for extrasolar planets in T-Type configuration we are 
interested in the ETV signal of the secondary star caused by an additional 
planet. 
This method is particularly important in the case when the planet's orbit 
may not be in the line of sight. 
However, such planets cause perturbations in the orbit of the transiting star,
leading to detectable ETVs. These were investigated for TTVs in several 
articles like e.g. \citet{mira05}; \citet{holman05} and \citet{agol07}.
The feasibility of the detection of extrasolar planets by 
the partial occultation on eclipsing binaries was investigated 
by~\citet{schneider90}.
Therefore, we set our goal to show which planet sizes for the T-Type 
configuration are detectable in the ETV signal of the secondary star with 
current observational equipment. In order to approximate the
detectability of possible extrasolar planets in T-type motion
by means of ETVs we used the work of \citet{syb10}. In principle they 
determined the sensitivity of the eclipse timing technique to circumbinary 
planets (exoplanets in P-Type motion) for space-based photometric observations.
 They showed that the typical photometric error (detectable timing 
amplitude dT) for CoRoT is about dT = 4 sec for a brightness (L) of 12~[mag] 
and dT = 16 sec for L=15.5~[mag]. 
Kepler has a dT = 0.5 sec for L=9~[mag] and a dT = 4 sec for L=14.5~[mag].
Future space missions will support the effort to detect new planets and 
smaller planets like for example: PLATO (Planetary Transits and Oscillations 
of stars) will monitor relatively nearby stars to hunt for Sun-Earth analogue 
systems \citep{rauer14}.
TESS (Transiting Exoplanet Survey Satellite) space mission is dedicated to 
detect nearby Earth or super-Earth-size planets on close-in orbits around the 
brightest M dwarfs~\citep{ricker14}. CHEOPS (Characterising ExOPlanets 
Satellite) will examine transiting exoplanets of known bright and nearby 
host stars~\citep{broeg13}. For our investigations we will use as detection 
criterion the photometric precision of CoRoT $dT_{crit}=16sec$.
\par
\begin{figure}
%\centerline
{\includegraphics[width=8.0cm,angle=0]{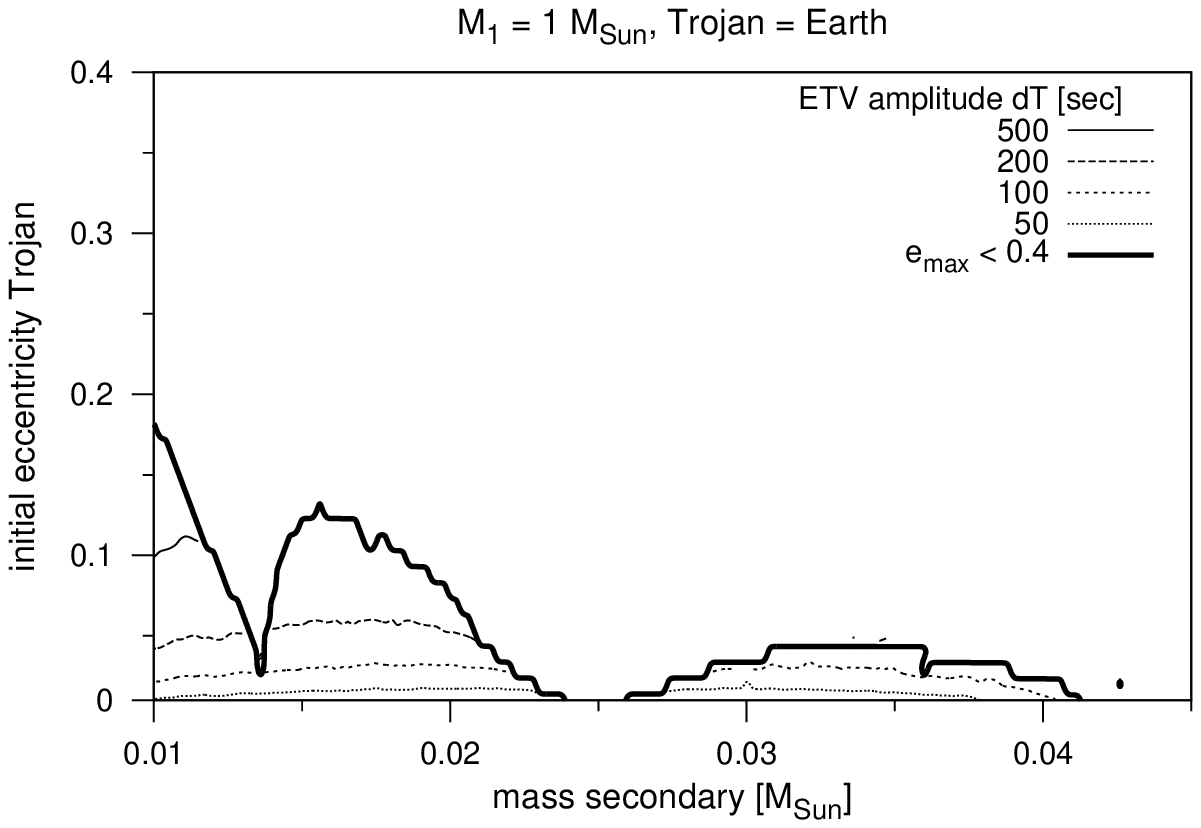}}
{\includegraphics[width=8.0cm,angle=0]{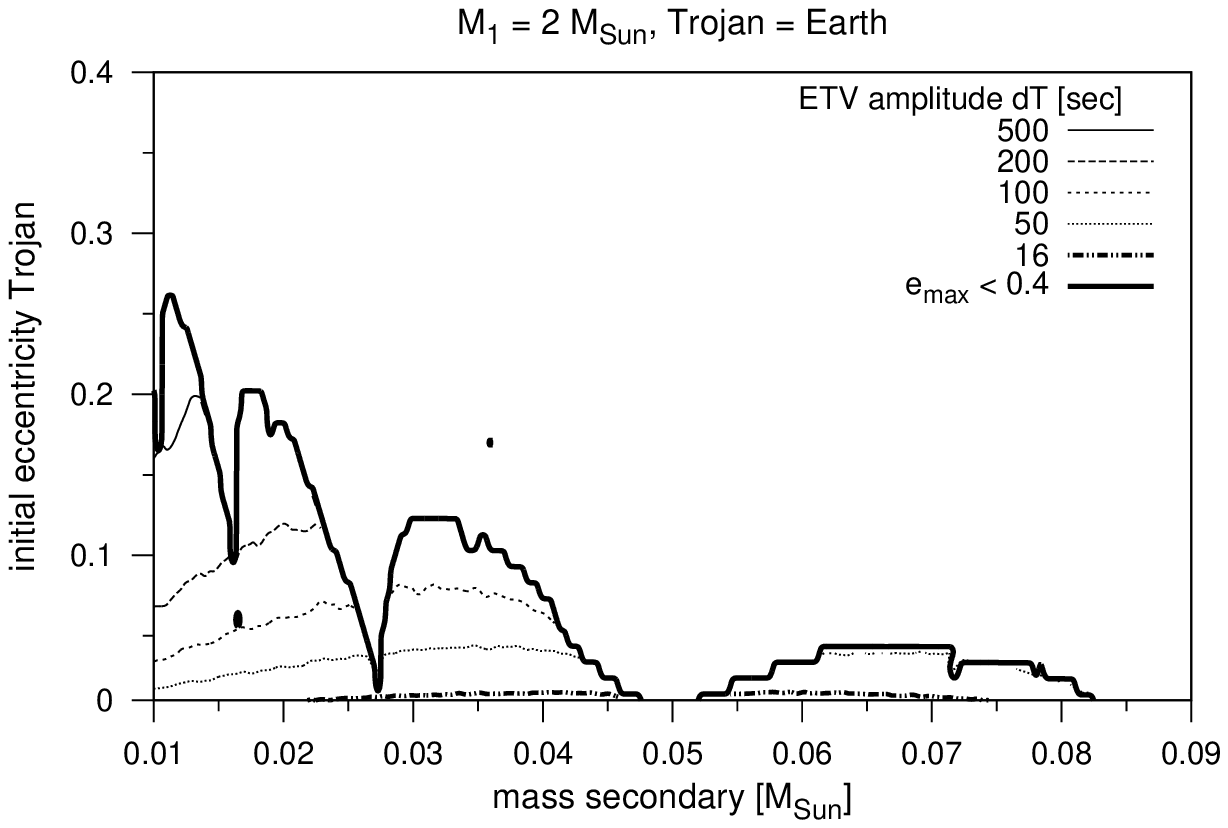}}
{\includegraphics[width=8.0cm,angle=0]{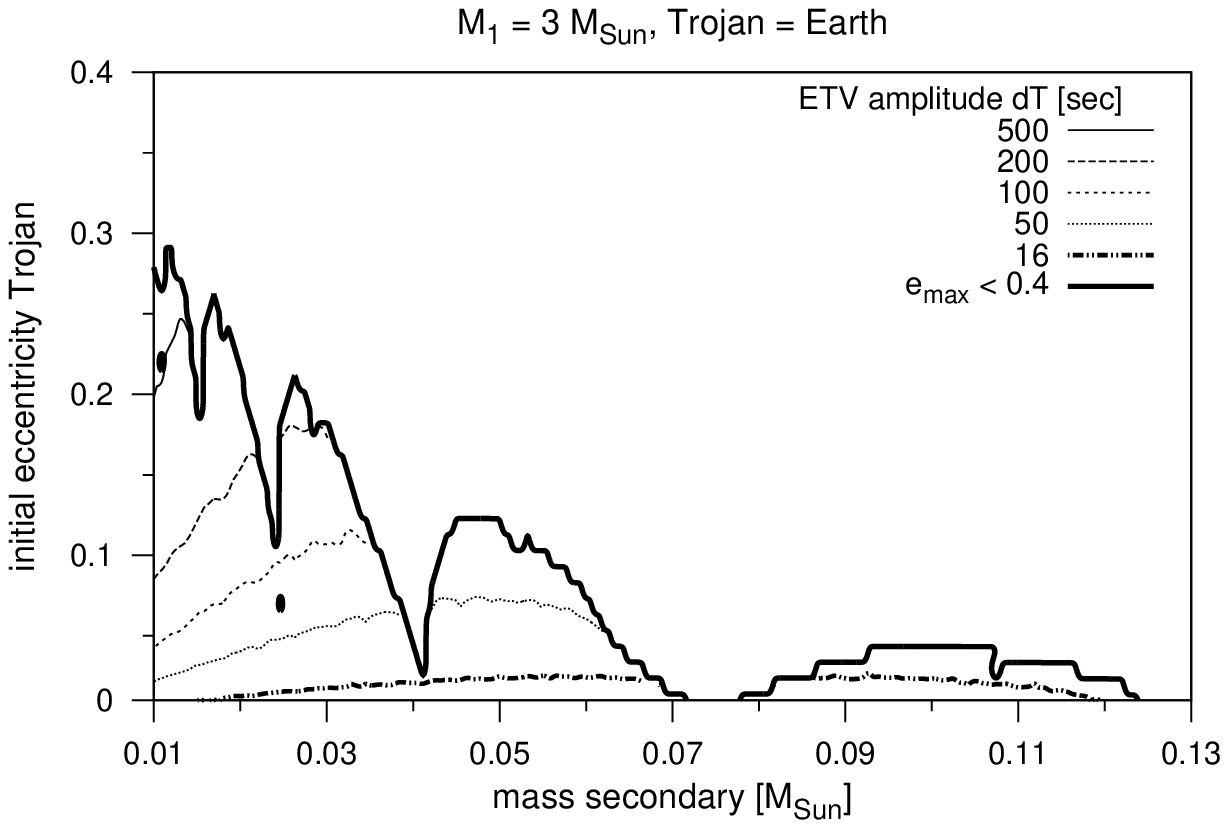}}
\caption{ETV map for $L_4$ in the planar ($i=0^{\circ}$) 
three-body problem for possible Trojan planets with one Earth mass, for 
1~$M_{sun}$ (upper graph) 2~$M_{sun}$ (middle graph), and  3~$M_{sun}$ 
(lower graph). The different lines presents the values of the amplitude of the 
ETV signal $dT$ in [s]. The outer (bold) line represent the stability border.}
\label{fig4}
\end{figure}
\begin{figure}
%\centerline
{\includegraphics[width=8.0cm,angle=0]{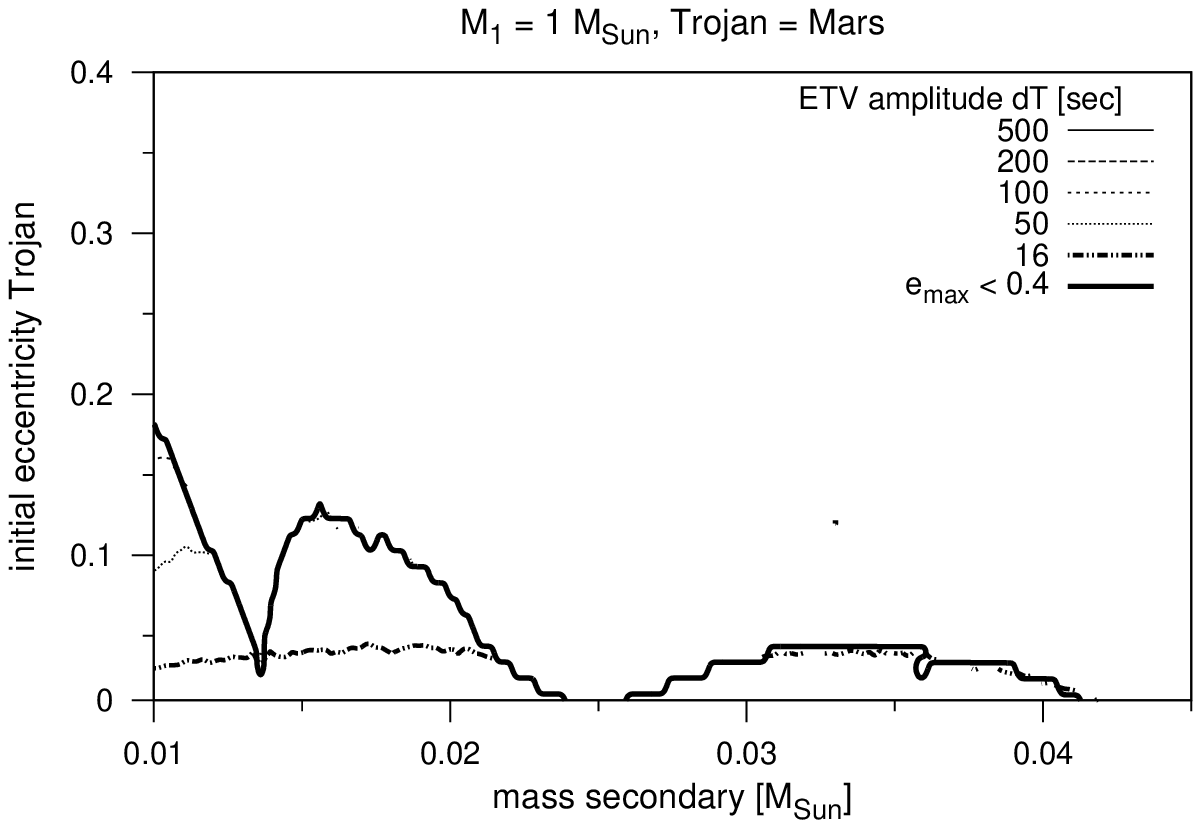}}
{\includegraphics[width=8.0cm,angle=0]{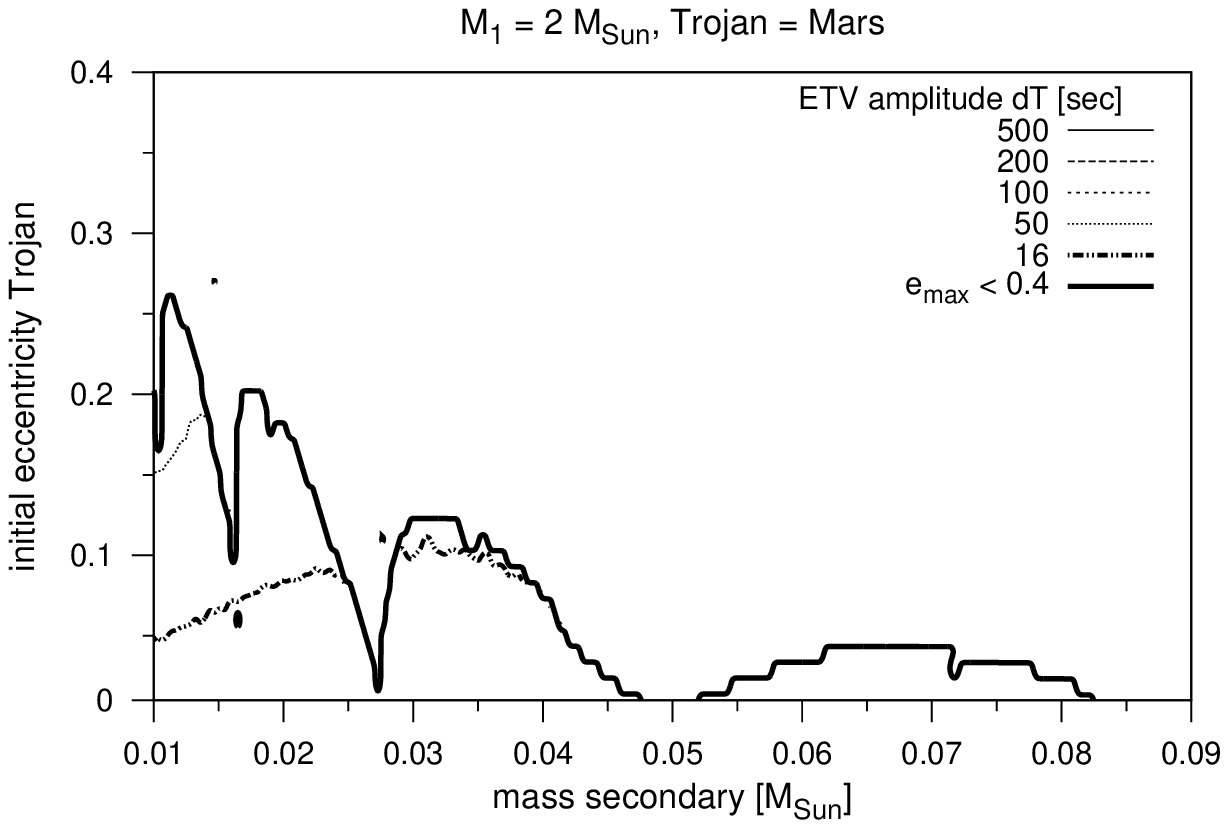}}
{\includegraphics[width=8.0cm,angle=0]{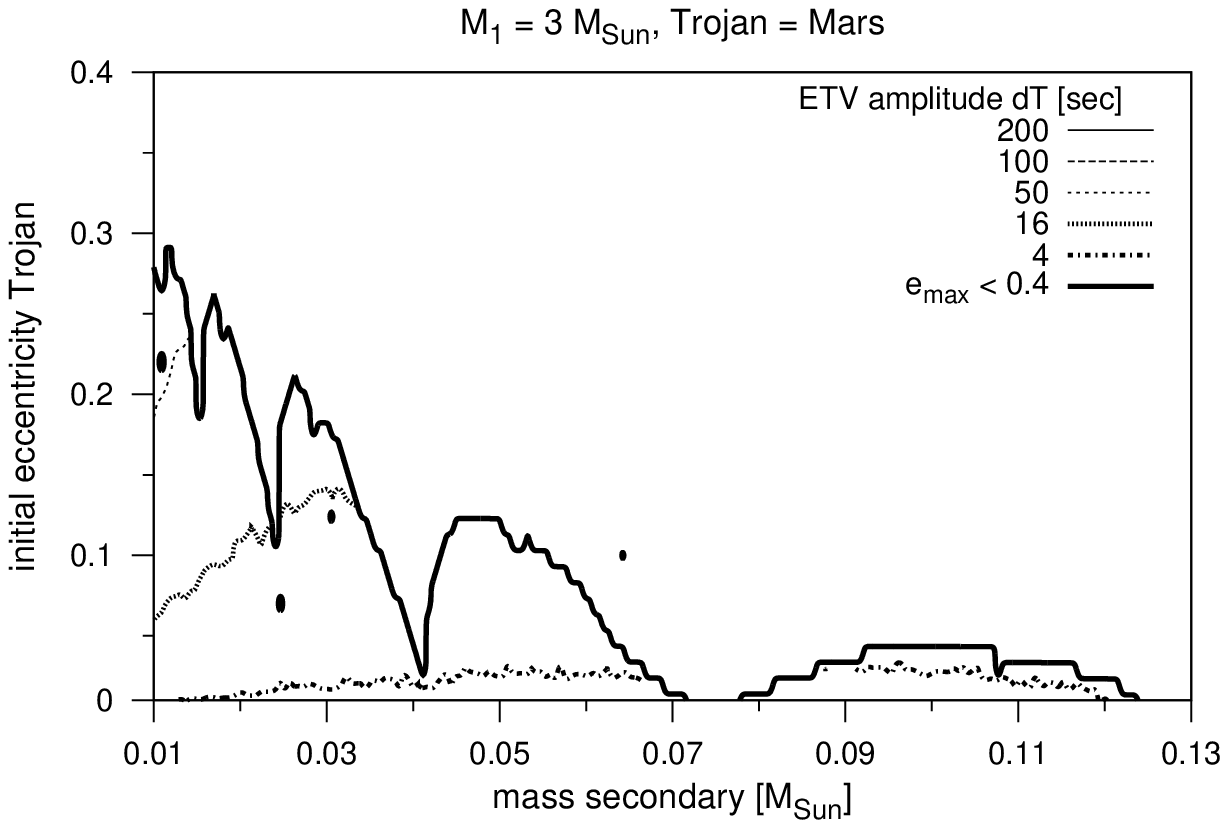}}
\caption{ETV map for $L_4$ in the planar ($i=0^{\circ}$) 
three-body problem for possible Trojan planets with one Mars mass, for 
1~$M_{sun}$ (upper graph) 2~$M_{sun}$ (middle graph), and  3~$M_{sun}$ (lower 
graph). The different lines presents the values of the amplitude of the ETV 
signal $dT$ in [s]. The outer (bold) line represents the stability border.}
\label{fig5}
\end{figure}

In our studies we investigated the amplitude of the ETV signals, 
therefore we varied the mass of the secondary body; starting with 
 a binary like system $m_2= 13 M_{Jup}$ (13 Jupiter masses corresponds to
the lowest mass of a brown dwarf). 
In addition, we varied the eccentricity of the Trojan planet whereas the 
secondary body moves in a circular orbit.
We used the stability analysis for the study of the ETV signals, which is 
presented in Fig.~\ref{fig2} for different masses of the Trojan body $M_{Jup}$, 
for 1~$M_{sun}$ (upper graph) 2~$M_{sun}$ (middle graph) and 3~$M_{sun}$ 
(lower graph). Whereas,  $m_1 \sim 1 M_{sun}$ represent Sun-like stars 
and $m_1 \ge$ 1 $M_{sun}$ depicts brighter stars. But we did not take into 
account less massive stars ($m_1$), because of the critical mass ratio $\mu_{crit}$, 
which means that the secondary star would be a planetary object instead of
a brown dwarf. 
As one can see the eccentricity of possible Trojan planets is limited 
by their eccentricity, that means Trojan planets can not have larger 
eccentricities than $e_{Tro}=0.2$ (for $m_2\geq 0.01$). This is also well 
visible in Fig.~\ref{fig4}. and Fig.~\ref{fig5} were we present contour
plots for terrestrial-like planets. Fig.~\ref{fig4} shows Trojans with
Earth mass and Fig.~\ref{fig5} shows Trojans with Mars mass. The most important
contours are the stability limit (outer border) and the border with 
$dT_{crit}=16s$. In case of an Earth-like Trojan planet with $m_1=M_{sun}$ 
all stable orbits produce a detectable ETV signal ($dT_{crit}\ge 16s$).
The figures~\ref{fig4} and \ref{fig5} showed that with larger mass of the 
central star ($m_1=2,3 M_{Sun}$) the stability limits increase, because of 
that Trojans can have larger eccentricities. 
However, the number of detectable ETV signals shrinks with larger $m_1$ and 
amplitude $dT$ have smaller values too, as 
concluded in Tab.~\ref{tab2}. This is mainly important for terrestrial like
Trojan planets. In case of $M_{Tro}$ is equal to Jupiter or Neptune mass
all stable orbits produce a detectable ETV signal. As represented 
in Tab.~\ref{tab2}, where the third column show the minimum amplitude
of the ETV signal $dT$ and fourth column depicts the number of detectable ETV 
signals within the stable region. We can summarize that possible terrestrial-like 
planets are detectable in binary star systems by the help of eclipse 
timing variations with restrictions.

\begin{table}
\centering
 \caption{Detectable timing amplitude $dT$ for different masses of the primary 
star (first column) and different type of the Trojan bodies are given in the 
second column starting with Mars mass ($m_{Tro}=0.032271\cdot10^{-6}M_{sun}$) 
then Earth mass ($m_{Tro}=0.30049\cdot10^{-5}M_{sun}$), Neptune mass 
($m_{Tro}=0.51684\cdot10^{-4}M_{sun}$), and Jupiter mass 
($m_{Tro}=0.9547\cdot10^{-3}M_{sun}$). The third column represents the minimum 
$dT$ and the last column depicts the percentage of all possible detectable 
signals ($dT_{crit}$) within the stable region, presented as bold line 
in Fig.~\ref{fig4} and Fig.~\ref{fig5}.}

\begin{tabular}{llll}
\hline
mass of  & type of   & minimum of & percent lie\\
the primary & Trojan & dT [h,m,s] & in the $dT_{crit}$\\
$[M_{sun}]$ & bodies &            &            \\
\hline
        1 & Mars     &  4.3s &   40 \\
        1 & Earth    & 24.2s &   100\\
        1 & Neptune  &  7.2m &   100\\
        1 & Jupiter  & 1.82h &   100\\
\hline
        2 & Mars     &  3.5s &   15\\
        2 & Earth    & 10.4s &   89\\
        2 & Neptune  &  2.1m &   100\\
        2 & Jupiter  & 0.82h &   100\\
\hline
        3 & Mars     &  2.6s &   3\\
        3 & Earth    &  6.0s &   77\\
        3 & Neptune  &  1.5m &   100\\
        3 & Jupiter  & 0.39h &   100\\
\hline
\end{tabular}
\label{tab2}
\end{table}

\section{List of candidates for binary (like) systems} 
As an application we prepared a list of possible candidates, which 
fulfil the most important dynamical and observational conditions.
In case of double stars it is very difficult to find candidates, because
the important parameters like semi-major axis $a$ and mass ratio $\mu$ of the
primary bodies are not or only partly given.
Maybe this problem will change in future, because of the data of the
space mission GAIA\footnote{Global Astrometric Interferometer for Astrophysics},
which will make a survey of one billion stars with many binary star systems~
\citet{eyer12} and binary-like systems~\citet{de12}.
For our search we used the 'Washington Visual Double Star Catalogue'~
\citep{mason01}. We found 24 systems with a mass ratio smaller than the 
critical one ($\mu \le 0.04$). But we found only one system where the 
semi-major axis from the secondary star $a$ is given, namely 
Antares ($\alpha$ Sco) with $a=2.9$~AU~\citep{worley} and $\mu \sim 0.016$.
Finally, we would like to mention that the distance of the secondary star should not be
larger than 1~AU, because of the long observation time span.
\par
At the moment we know about 1900 exoplanets in more than 1200 
planetary systems, among them 93 planets have masses larger then 10 Jupiter 
masses. We classified them as binary like systems, because these objects are 
brown dwarfs.
When we look at mass ratios which fulfil the criterion that 
$\mu \leq \mu_{crit}$, the list of candidates decreases to 68 candidates. 
Concerning to the condition that the distance to the central star should be 
not too far (that means $a \leq 1$ $AU$ because of long observation time) 
we found only 10 candidates.
The data of all planets are taken from the Exoplanet-catalogue maintained by J. 
Schneider.

\begin{table}
\centering
 \caption{List of candidates to detect possible Trojan planets in
binary-like systems. The list is sorted according to the semi-major axis 
($a_2$) of the brown dwarf. All systems were detected by TTVs except for those 
marked with a star (which were detected by radial velocity).}
  \begin{tabular}{lllll}
\hline  
 Name       &    mass    & $a$ in [AU]  & $m_1$       & $\mu \le \mu_{crit}$\\
            & [$M_{Jup}$]&              & [$M_{Sun}$]  &          \\
\hline
WASP-18 b   &    10.43   &     0.020            & 1.24    &   0.00796\\
KELT-1 b    &    27.38   &     0.025            & 1.335   &   0.01919\\
XO-3 b      &    11.79   &     0.045            & 1.41    &   0.00791\\
CoRoT-27 b  &    10.39   &     0.048            & 1.05    &   0.00935\\
CoRoT-3 b   &    21.77   &     0.057            & 1.41    &   0.01452\\
HD 162020 b* &    14.4    &     0.074            & 0.75    &   0.01799\\
Kepler-39 b &    18      &     0.155            & 1.1     &   0.01537\\
Kepler-27 c &    13.8    &     0.191            & 0.65    &   0.01985\\
HD 114762 b*&    10.98   &     0.353            & 0.84    &   0.01232\\
HD 202206 b*&    17.4    &     0.830            & 1.13    &   0.01448\\
  \hline
\end{tabular}
\label{tab3}
\end{table}

\section{Conclusions}
We have carried out a parameter study for a variety of binaries with 
different mass-ratios of the primary bodies and different orbital elements of 
the Trojan planet. 
The stability analysis has shown that there exist a stable region for planets 
-- gas giants as well as terrestrial like planets -- in T-type motion for low 
eccentricities $e_{Tro} \le 0.25$.
Our results have shown that binary-like systems (the secondary body is a 
brown dwarf) and binary systems with low-mass stars are good candidates to 
discover new planets by eclipse timing of the secondary star, agreeing with 
the work of \citet{schneider95}.
At this point we have to concede that from an observational point of view 
M-stars are better candidates than brown dwarfs.
Nevertheless, ETV signals of brown dwarfs would be detectable too.
We found that for all stable initial conditions an Earth-like Trojan
planet would produce detectable ETV signals if the primary star has 1 $M_{sun}$.
This is also valid for Jupiter and Neptune type Trojan planets, but 
also for not Sun-like stars ($m_1= 2M_{sun}$ and $m_1= 3M_{sun}$).
Unexpectedly, we found that even Trojan planets with Mars mass would be detectable, 
but under favourable circumstances. 
As an application we prepared a list of possible candidates for the 
detection possible Trojan planets of binary and binary-like systems. In case of binaries 
we found 24 candidates, but for only a single system (Antares, $\alpha$ Sco) the semi-major 
axis $a$ is constrained. In addition, we made a list of candidates for binary-like stars 
which is given in Tab.~\ref{tab3} where we found 10 candidates. 
\par
We can conclude that possible terrestrial-like planets are detectable in 
binary star systems by the help of eclipse timing variations with 
restrictions. However, future space missions will have a better photometric 
precision which will enlarge the number of detectable ETV signals.

\section*{Acknowledgments}
R. Schwarz, B. Funk and \'A. Bazs\'o want to acknowledge the support by
the Austrian FWF project P23810-N16.

%\appendix

%\section[]{Large gaps}

%(This appendix was not part of the original paper by
%A.V.~Raveendran and is included here just for illustrative
%purposes. The references are not relevant to the text of the
%appendix, they are references from the bibliography used to
%illustrate text before and after citations.)

\end{document}